# Defending *The Fallacy of Fine-Tuning*

Victor J. Stenger

University of Colorado, Boulder, Colorado

University of Hawaii, Honolulu, HI

January 28, 2012

## Abstract

In 2011, I published a popular-level book, *The Fallacy of Fine-Tuning: Why the Universe is Not Designed for Us*. It investigated a common claim found in contemporary religious literature that the parameters of physics and cosmology are so delicately balanced, so "fine-tuned," that any slight change and life in the universe would have been impossible. I concluded that while the precise form of life we find on Earth would not exist with slight changes in these parameters, some form of life could have evolved over a parameter range that is not infinitesimal, as often claimed. Postdoctoral fellow Luke Barnes has written a lengthy, highly technical review of the scientific literature on the fine-tuning problem. I have no significant disagreement with that literature and no prominent physicist or cosmologist has disputed my basic conclusions. Barnes does not invalidate these conclusions and misunderstands and misrepresents much of what is in the book.

## 1. Introduction

In 2011, I published a popular-level book, *The Fallacy of Fine-Tuning: Why the Universe is Not Designed for Us*.[1] It investigated a common claim found in contemporary religious literature that the parameters of physics and cosmology are so delicately balanced, so "fine-tuned," that any slight change and life in the universe would have been impossible. I concluded that while the precise form of life we find on Earth would not exist with slight changes in these parameters, some form of life could have evolved over a parameter range that is not infinitesimal, as often claimed.

The simplest solution to the fine-tuning problem, and the favorite among scientific experts, is that our universe is just one in a multitude of universes and we just happen to live in the one suited for us. While I fully respect this possibility, I have limited my investigation to a single universe.

Postdoctoral fellow Luke Barnes has written a lengthy, highly technical review of the scientific literature concerning the fine-tuning problem titled "The Fine-Tuning of the Universe for Intelligent Life"[2] *The Fallacy of Fine-Tuning* did not address the scientific literature. Barnes' paper is written for experts in the field, who were not my intended audience and with whom I have no significant scientific disagreements. Barnes does not challenge my basic conclusions. Nor, to my knowledge, has anyone on the long list of reputable physicists and cosmologists who Barnes insists believe in fine-tuning. In fact, several were consulted in writing the book.

*Fallacy* was concerned with the widespread argument found in theological and religious apologetic writings that the putative fine-tuning of the parameters of physics and cosmology cannot be the product of purely natural forces.[3] I agree that life, as we know it on Earth, would not exist with a slight change in these parameters. However, there is no reason to limit ourselves to earthly life but consider the possibility of other forms of life, carbon-based or otherwise.

Depending on what you count, about thirty parameters are generally suggested as being fine-tuned. Of these, some theists have claimed that five parameters exist that are so exquisitely fine-tuned that changing any single one by one part in $10^{40}$ or more would mean that *no* life of any kind was possible. These crucial parameters are:

1. The ratio of electrons to protons in the universe
2. The expansion rate of the universe
3. The mass density of the universe
4. The ratio of the electromagnetic and gravitational forces
5. The cosmological constant

In *Fallacy*, I give plausible reasons for the values of each within existing, well-established physics and cosmology.

The remaining parameters are also supposed to be fine-tuned to many orders of magnitude. I show that they are at best fine-tuned, if you want to call it that, to 10-20 percent. Barnes seems to want me to reduce this to maybe 1-5 percent. But nowhere does he show that they should be $10^{-40}$. My essential point is, when all parameters are taken together the region of parameter space that should allow some form of life to evolve is not the infinitesimal point that the theist literature would want us to believe.

In *Fallacy*, I formulate some of my arguments with certain simplified assumptions, such as semi-Newtonian cosmology. Barnes attacks these by using higher-level arguments that are quite irrelevant. He fails to explain why my simplifications are inadequate for my purposes.

In short, Barnes objections are largely superfluous. However, I cannot leave it at that since he has in several places misrepresented and misunderstood what I have said. In what follows, I will attempt to clarify these issues.

## 2. Point-of-View Invariance (Chapter 4)

Barnes writes, "Are the laws of nature themselves fine-tuned? Stenger defends the ambitious claim that the laws of nature could not have been different because they can be derived from the requirement that they be *Point-of-View Invariant* (hereafter, PoVI)."

He continues, "We can formulate Stenger's argument for this conclusion as follows:

LN1. If our formulation of the laws of nature is to be objective, it must be PoVI.
LN2. Invariance implies conserved quantities (Noether's theorem).
LN3. Thus, "when our models do not depend on a particular point or direction in space or a particular moment in time, then those models must necessarily contain the quantities linear momentum, angular momentum, and energy, all of which are conserved. Physicists have no choice in the matter, or else their models will be subjective, that is, will give uselessly different results for every different point of view. And so, the conservation principles are not laws built into the universe or handed down by a deity to govern the behavior of matter. They are principles governing the behavior of physicists."

Barnes continues, "This argument commits the fallacy of equivocation—the term 'invariant' has changed its meaning between LN1 and LN2. The difference is decisive but rather subtle, owing to the different contexts in which the term can be used. We will tease the two meanings apart by defining covariance and symmetry, considering a number of test cases." He follows with a lengthy example:

**"Galileo's Ship:** We can see where Stenger's argument has gone wrong with a simple example, before discussing technicalities in later sections. Consider this delightful passage from Galileo regarding the brand of relativity that bears his name:

> Shut yourself up with some friend in the main cabin below decks on some large ship, and have with you there some flies, butterflies, and other small flying animals. Have a large bowl of water with some fish in it; hang up a bottle that empties drop by drop into a wide vessel beneath it. With the ship standing still, observe carefully how the little animals fly with equal speed to all sides of the cabin. The fish swim indifferently in all directions; the drops fall into the vessel beneath; and, in throwing something to your friend, you need throw it no more strongly in one direction than another, the distances being equal; jumping with your feet together, you pass equal spaces in every direction. When you have observed all these things carefully (though doubtless when the ship is standing still everything must happen in this way), have the ship proceed with any speed you like, so long as the motion is uniform and not fluctuating this way and that. You will discover not the least change in all the effects named, nor could you tell from any of them whether the ship was moving or standing still.

"Note carefully what Galileo is not saying. He is not saying that the situation can be viewed from a variety of different viewpoints and it looks the same. He is not saying that we can describe flight-paths of the butterflies using a coordinate system with any origin, orientation or velocity relative to the ship. Rather,

Galileo's observation is much more remarkable. He is stating that the two situations, the stationary ship and moving ship, which are externally distinct are nevertheless internally indistinguishable. "

Barnes carries on for another page with this lesson from Galileo. But I need not quote it any further, because I don't have any quarrel with Galileo. Barnes has grossly misrepresented me if he is claiming that I have said that "the situation can be viewed from a variety of different viewpoints and it looks the same." In fact, he quotes exactly what I did say: "The models of physics cannot depend on the point of view of the observer." These statements are not equivalent.

Of course, different observers see different things. Suppose someone drops a rock from the top of the mast of Galileo's moving ship. An observer on the ship will see it fall in a straight line, while an observer on shore will see it fall along a parabolic path. My point is that these paths cannot be built into the models of physics since they depend on point-of-view. And, by applying this principle, the working models of physics do not have these different paths built in. In freshmen physics, students learn to calculate projectile paths from the same set of equations and apply them in different frames of reference.

Also, I have never claimed that *all* the laws of physics follow from point-of-view invariance. PoVI is a necessary principle, but it does not by itself determine all the laws of physics. There are choices of what transformations are considered and any models developed must be tested against the data. However, it is well established, and certainly not my creation, that conservation principles and much more follow from symmetry principles. Other principles can be connected to spontaneously broken symmetries. I include gauge symmetry as an application of PoVI. The notion here is that physicists are not completely free to invent any model they want when discussing this or any hypothetical other universe. For example, if they are to maintain the notion that there is no special point in space, then they can't suggest a model that violates momentum conservation.

Barnes quotes my statement, "Physicists are forced to make their models Lorentz invariant so they do not depend on the particular point of view of one reference frame moving with respect to another." (p. 82) He says, "This claim is false. Physicists are perfectly free to postulate theories which are not Lorentz invariant, and a great deal of experimental and theoretical effort has been expended to this end."

Of course, physicists are free to postulate all the theories they want. But no physicist is going to propose a model that depends on his location and his point of view. Events that depend on time and place are discrete incidences like history and geography, not the universal processes described by physics. Quite simply, much of existing, empirically verified physics follows from a principle in which physicists force themselves to construct their models to be independent of the observer's point of view. If, someday, experiment shows a violation of this principle, then we will have to discard it. So far, this has not happened, as Barnes points out. However, the standard model of elementary particles does include the breaking of gauge symmetry at low energies in order to describe observations from that special point of view. The underlying principles of the model, however, remain point-of-view invariant.

Barnes objects to my association of gauge invariance with PoVI, but gives no reason. Instead, he quotes various authors to the effect that gauge invariance could be wrong. Of course, it could be wrong. In all my books I emphasize that I am perfectly at home with all scientists who are ready to change their ideas the moment the data require them to do so.

## 3. Gravity is "Fiction" (Chapter 7)

Barnes disagrees with my referring to gravity as a "fictitious" force. We call the centrifugal and Coriolis forces "fictitious" because we can find a reference frame in which they are not observed. I point out the same is true for the gravitational force. An observer in a falling capsule, such as a spacecraft in orbit, experiences no gravitational force.

The disagreement here is over our differing philosophical views on the nature of physics. Barnes is a *Platonic realist* who considers the laws of physics as inherent ingredients of reality. I am a *commonsense realist* who holds that the so-called "laws of physics" are simply the human-invented ingredients of models that physicists introduce to describe observations. They are all fictitious, as far as I am concerned, and while they must agree with the data, we have no way of knowing precisely what they have to do with reality.

In his theory of general relativity, Einstein replaced the gravitational force with paths along geodesics in curved space-time. That is, there is no gravitational force in general relativity. Certainly, gravity is a real phenomenon. However, the gravitational force is fiction. In this and most of Barnes' other comments, we don't disagree on the physics so much as how to characterize and interpret it.

## 4. Entropy at the Beginning of the Universe (pp. 107-113)

Barnes says, "Stenger's assertion that 'the universe starts out with maximum entropy or complete disorder' is false. A homogeneous, isotropic spacetime is an incredibly low entropy state."

Here Barnes fails to grasp the argument being made, that a volume of space can have maximal entropy and still contain very low entropy as compared to the visible universe. Assume our universe starts out at the Planck time as a sphere of Planck dimensions. Its entropy will be as low as it can be. However, at the same time, a Planck sphere is akin to a black hole whose entropy is maximal for an object of the same radius. It is not logically inconsistent to be both low and maximum at the same time.

In short, the universe could have started out in complete disorder and still produced organized structures. The reason is, as the universe expands its maximum allowed entropy grows with it so that order can form without violating the second law of thermodynamics.

## 5. Carbon and Oxygen Synthesis In Stars (Chapter 9)

Barnes says that I have failed to "turn back the force" of the claim that the parameters of physics are fine-tuned to allow carbon and oxygen to be synthesized in stars. However, he quotes Weinberg (as I do) as saying that this phenomenon "does not seem to me to provide any evidence for fine tuning." The best Barnes can do is to refer to some additional studies that are what he calls "highly suggestive" that carbon and oxygen production would be "drastically curtailed by a tiny change in the fundamental constants."

I have shown, based on published calculations of others, that carbon and oxygen synthesis do not, as Weinberg says, "provide any evidence for fine tuning."

## 6. Expansion Rate of the Universe

The energy density and expansion rate of the universe are two of five parameters of the universe that, as mentioned, are said to be fine-tuned to over forty orders of magnitude.

Theistic fine-tuners, such as William Lane Craig[4] and Dinesh D'Souza,[5] often quote Stephen Hawking out of context in this regard. On page 121 of his 1988 blockbuster bestseller, *A Brief History of Time,* Hawking said "If the rate of expansion one second after the big bang had been smaller by even one part in a hundred thousand million million, the universe would have recollapsed before it ever reached its present size."[6]

The fine-tuners fail to mention that a few pages later, on page 128 of *Brief History*, Hawking said, "The rate of expansion of the universe would automatically become very close to the critical rate determined by the energy density of the universe. This could then explain why the rate of expansion is still so close to the critical rate, without having to assume that the initial rate of expansion of the universe was very carefully chosen."[7]

The expansion rate and energy density are not independent parameters. In *Fallacy,* I provided the equations that demonstrate this, showing that neither is fine-tuned for life. (Chapter 5).

Barnes does not challenge this essential point, but rather goes into detail on the problems of inflation, showing that it could be wrong. Of course, but again I am limiting myself to existing knowledge and so far inflationary cosmology has not been falsified and helps explain many observations. Here I simply reiterate the point made by Hawking in 1988 that inflation could account for the fact that the expansion rate seems to be fine-tuned.

## 7. Gravity and the Masses of Particles (Chapter 7)

Barnes similarly misrepresents the case I make against one of the most common, fine-tuning claims, that gravity is 39 orders of magnitude weaker than electromagnetism, and, if this were not so, we would not exist. I point out the elementary physics fact that this is only true for a proton and electron. In general, the relative strength of the two forces depends on the masses and charges of the particles involved. I explain that the reason gravity is so much weaker than electromagnetism for elementary particles is because of their low mass compared to the Planck mass. I then propose a plausible explanation for this low mass, namely, in the standard model the masses are intrinsically zero and their observed masses are the result of small corrections, such as the Higgs mechanism.

Barnes reacts, "Stenger is either not aware of the hierarchy and flavour problems, or else he has solved some of the most pressing problems in particle physics and not bothered to pass this information on to his colleagues." So, Barnes will not accept my argument until I solve the hierarchy and flavor problems, certainly a daunting task. But I claim I don't have to. I just have to suggest a plausible reason, consistent with our best existing knowledge, why the masses of particles are small. As long as no one can disprove this explanation, I win the argument.

Similarly, I give plausible reasons for the mass differences of protons, neutrons, and electrons. Barnes again misrepresents me, claiming that my statement that "the mass difference between the neutron and proton results from the mass difference between the d and u quarks" is "false, as there is also a contribution from the electromagnetic force." He ignores the fact that I explicitly attribute the mass differences of the d and u quarks to the electromagnetic force (*Fallacy* p. 178).

## 8. Strengths of Forces

On page 189 of *Fallacy* I said, "All the claims of the fine-tuning of the forces of nature have referred to the values of the force strengths in our current universe. They are assumed to be constants, but, according to established theory (even without supersymmetry), they vary with energy." Barnes says the first sentence is "false by definition—a fine-tuning claim necessarily considers different values of the physical parameters of our universe." Once again, he is accusing me of saying something I did not say. I did not say different values are not "considered." Of course they are. The point is that all the studies I have looked at (remember, I focus on the theistic literature) treat these force parameters as constants, when they are not.

Barnes also fails to grasp the point I make that the force constants are thought to be related to one another and expected to come together at some high unification energy (See Fig. 10.4, p. 189). The fact that they only differ now by a factor of six should not be regarded as fine-tuning.

Barnes says, "to show (or conjecture) that a parameter is derived rather than fundamental does not mean that it is not fine-tuned." Right. And the fact that we can't prove that Bertrand Russell's teapot is not orbiting the sun between Mars and Jupiter does not mean it is.

## 9. Charge Neutrality

Another fine-tuning claim is the ratio of protons to electrons in the universe (p. 205). I argue that this parameter results from charge conservation and quote from a book on cosmology written by an astronomer. Barnes agrees that this is not a good fine-tuning argument, but objects to my explanation, it seems, because I relate it to PoVI. He says, "Charge conservation follows from gauge invariance, but gauge invariance does not follow from 'point of view invariance' as Stenger claims."

In my 2006 book *The Comprehensible Cosmos*, I argued that gauge invariance is a form of point-of-view invariance.[8] Barnes disagrees, but again his disagreeableness does not change the conclusion here.

## 10. MonkeyGod (Chapter 13)

Barnes finds much fault with my simple program *MonkeyGod*, which I put on my website years ago to allow people to "create their own universe." I included a description of it in *Fallacy* so that readers could see exactly what the program does. I clearly called these "toy universes" (p. 236), but figured they were still useful for giving us some feel for the dependence of certain quantities on the basic parameters of physics. One quantity of significant relevance to the fine-tuning question is stellar lifetime. I claim no deep results, but find it interesting that a wide range of fundamental physics constants will give long lifetime stars, a likely prerequisite for life.

Barnes makes his usual objections to my admitted oversimplifications. Does he really expect me to simulate entire universes?

## 11. Conclusion

Barnes objections to *The Fallacy of Fine-Tuning* result from a misunderstanding of my intention in writing the book, and both a misunderstanding and misrepresentation of much that is in it. My intent was to investigate the claim found in much theistic literature that carbon-based life, as we know it, would be impossible if any one of thirty or so parameters of physics and cosmology changed by an infinitesimal amount. Five of these are critical parameters for which it is claimed no form of life would be possible without the posited fine-tuning.

I have never denied that life, as we know it on Earth, would not have evolved with slight changes in parameters. In *Fallacy* I showed (1) that plausible explanations, consistent with existing knowledge, can be made for the observed values of the five critical parameters and, (2) plausible ranges for the other parameters exist that are far from infinitesimal, contrary to what is claimed in the theistic literature.

Nothing in Barnes' paper changes my basic conclusion: The universe is not fine-tuned for us. We are fine-tuned to the universe.

## Acknowledgements

Many thanks to Raymond Briggs, Kim Clark, Jonathan Colvin, Yonatan Fishman, Craig James, Bill Jefferys, John Kole, Don McGee, Brent Meeker, and Bob Zannelli for helping me prepare this paper.